\documentclass[prb,preprint,showpacs,preprintnumbers,amsmath,amssymb]{revtex4}
\usepackage{graphicx}% Include figure files

\usepackage{dcolumn}% Align table columns on decimal point
\usepackage{bm}% bold math
\usepackage[mathscr]{eucal}
\usepackage{mathrsfs}

\begin{document}

% Full title of the paper (Capitalized)
\title{First-principles study on the specific heat of glass at glass transition 
with a case study on glycerol
}
% \title{First-principles calculation of the specific-heat jump in glass transition: a case study  on glycerol}
% \title{A thermodynamics investigation on glass transition}

% Authors, for the paper (add full first names)
\author{Koun Shirai, Kota Watanabe, and Hiroyoshi Momida}
% \email{koun@sanken.osaka-u.ac.jp}
\affiliation{%
The Institute of Scientific and Industrial Research, Osaka University, 8-1 Mihogaoka, Ibaraki, Osaka 567-0047, Japan
}%

% Abstract (Do xnot use inserted blank lines, i.e. \\) 
\begin{abstract}
The standard method to determine the transition temperature ($T_{g}$) of glass transition is the jump in the specific heat $\Delta C_{p}$. Despite this importance, standard theory for this jump is lacking. The difficulties encompass from lack of proper treatments of specific heat of liquids, hysteresis, to the timescale issue. 
The first part of this paper provides a non-empirical method to calculate specific heat in the glass transition, with resolving these difficulties. The method consists of molecular dynamics (MD) simulations based on density-functional theory (DFT) and thermodynamics methods. The total-energy approach based on DFT, in which the total energy is the most reliable energy for any state of matter, enables us to calculate specific heat, irrespective of solids or liquids. A serious problem for glass-transition states is involvement of complicated energy dissipation processes. This problem is resolved by employing adiabatic MD simulations, by which the relationship between the internal energy and equilibrium temperature is calculated. The problems of hysteresis and the timescale issue are alleviated by restricting the scope of calculations to equilibrium states only.
The second part of this paper describes an application of the theory to the specific-heat jump of glycerol in order to show the validity of the methods. In spite of severe limitations due to the small size of supercells, a reasonable value for the specific-heat jump is obtained. By decomposing $\Delta C_{p}$ into contributions of the structural energy, phonon, and thermal expansion parts, we have a sound interpretation for the specific-heat jump: the major contribution to $\Delta C_{p}$ comes from the change in the structural energy. From this, a neat energy diagram about the glass transition is obtained: this greatly help our understanding on the glass transition. 
% $\{ \bar{\bf R}_{j} \}$, but neither from the phonons nor from thermal expansion. 
An outcome of this study is verification of the empirical relationship between the fragility and specific-heat jump, each reflecting the change in the energy barrier and the change in the internal energy, respectively. These two energies are scaled by the ratio $k=T_{g}/\Delta T_{g}$, where $\Delta T_{g}$ is the width of the transition, through which the two quantities are interrelated. This is useful for organizing various relationships that are found empirically.
% This gives a firm ground to the above empirical relationship.
\end{abstract}

\pacs{Rev. 1.0: 31 Jan 2022}
% deflect from

\maketitle
%%%%%%%%%%%%%%%%%%%%%%%%%%%%%%%%%%%%%%%%%%
%% Only for the journal Gels: Please place the Experimental Section after the Conclusions

%%%%%%%%%%%%%%%%%%%%%%%%%%%%%%%%%%%%%%%%%%

\section{Introduction}
\label{sec:intro}
Specific-heat measurement is an elemental method for studying the glass transition.\cite{Davies53a,Nemilov-VitreousState,Rao02}
The jump $\Delta C_{p}$ in specific heat when scanning temperature $T$ is a signature of the glass transition: here $\Delta C_{p}=C_{p}^{(l)}-C_{p}^{(g)}$, and $C_{p}^{(g)}$ and $C_{p}^{(l)}$ are isobaric specific heat of glass and supercooled liquid, respectively. In this paper, when a simple word ``liquid" is used, it denotes the supercooled liquid, because here we study the liquid state only at temperatures lower than the melting temperature $T_{m}$. When the liquid state at $T>T_{m}$ is inferred, it is called the ``normal" liquid.
The temperature at which the jump occurs defines the glass-transition temperature $T_{g}$. The glass transition occurs in a finite temperature range from $T_{g1}$ to $T_{g2}$ with the width $\Delta T_{g}=T_{g2}-T_{g1}$. This temperature range is called the {\em transition region}.\cite{Moynihan74} At this temperature range, other thermodynamic properties also exhibit similar jumps: volume $\Delta V$, compressibility $\Delta \kappa$, thermal expansivity $\Delta \alpha$, etc.\cite{Kauzmann48}
Among these jumps at $T_{g}$, a relationship exists which is known as the Prigogin-Defay relation.\cite{Prigogine54} Although the Prigogin-Defay relation has been extensively studied since early days,\cite{Davies53a} debate is continued as yet.\cite{Moynihan81,Nieuwenhuizen97,Schmelzer06}
Another interest is the nature of the excess entropy $S_{\rm ex}$ of supercooled liquids, which is directly obtained by the jump $\Delta C_{p}$. The excess entropy is considered as the configurational entropy $S_{c}$ of supercooled liquid, which plays a fundamental role in glass behaviors.\cite{Tatsumi12,Dyre18,Berthier19a} In connection to the Adam-Gibbs theory \cite{Adam65}, the relation between kinetic and thermodynamic fragilities are often discussed.\cite{Angell76, Takahara95,Ngai99,Johari00,Martinez01}
Angell pointed out that there is a relation between fragility and $\Delta C_{p}$: fragile glasses have large $\Delta C_{p}$ while strong glasses small one.\cite{Angell95} Some explanations for this relation based on models are proposed.\cite{Granato02,Xia00,Lubchenko07, Garrahan03,Biroli05,Chandler10}
Although each of them may capture some of glass properties from their own viewpoints, it is unclear whether these theories are equivalent or not, because they are described in different terms. It is desirable to calculate $\Delta C_{p}$ without models.

In spite of above important roles that the specific-heat jump plays in the glass research, little study has been reported to calculate $\Delta C_{p}$ itself. 
There are multiple reasons for this. First, although the method of calculation of specific heat is well established for solids, this is lacked for liquids:\cite{note-liquid} a brief review on model calculation for specific heat of liquids is given by Granato.\cite{Granato02}
Present theories of $\Delta C_{p}$ depend on the models for liquids.\cite{Trachenko11,Garden12}
Second, there seems no unique relation between $C_{p}$ and $T$ in the glass transition because of hysteresis: the specific heat during the transition has dependence on the cooling/heating rate.\cite{Moynihan76,Hodge83} 
For this reason, the majority of studies focus the kinetic aspect of the specific heat by taking the relative change, $\tilde{C}_{p}(T) = (C_{p}(T)-C_{p}^{(g)})/\Delta C_{p}$, only with discussing the relaxation effect of glass transition.\cite{Hodge94} The third is a serious obstacle when specific heat is calculated by molecular-dynamics (MD) simulations. This study is the case. The timescale of the glass transition is of a macroscopic scale (longer than 1 s), which is by far longer than the simulation time that present MD simulations can perform. This casts doubt that the achievements of even classical MD simulations are disparate from the real transition.

The first problem can be solved by using first-principles (FP) calculations based on density-functional theory (DFT), which is the today's most reliable method for calculating properties of solids. The total energy of materials is accurately obtained for a given structure, regardless of equilibrium or nonequilibrium states. 
The second problem can be avoided by restricting our study to equilibrium states only. The rate dependence is caused by the atom relaxation, that is, the retardation in the atom rearrangements for the external influence. Here, we are concerned with the final state of equilibrium only. This means that the calculated $C_{p}$ at a $T$ represents the limiting value obtained by as slow the cooling/heating rate as possible.
The third problem is managed again by restricting our study to equilibrium states only. This is contrast to other studies employing MD simulations, where the time evolution of glass material is pursued. 
Details of these explanations are given later. 

In this paper, the jump in specific heat of glasses is studied by combination of FP MD simulation and thermodynamic methods. On account of the above difficulties in calculating $\Delta C_{p}$, in the first part of this paper, general principles for calculating specific heat of liquids and glasses are given, which is described in Sec~\ref{sec:Theory}. Recently, essentially the same approach based on the total-energy theory has been published by Han et al.\cite{Han20} Their approach is based on the entropic representation. Here, the energy representation with explicitly specifying state variables is employed. This has a great merit in providing useful insights into the nature of the glass transition.
The second part gives a working example of the present method, which is described in Sec~\ref{sec:Application}. As a concrete material, glycerol is studied. The reasons of this choice are that, first, glycerol often serves a  material as benchmark of study: it was studied at an early stage of glass research \cite{Gibson23,Oblad37}, and it continues to offer a new front of research \cite{Birge85,Massalska-Arodz91,Kojima93,Christensen94}. Second, the specific-heat jump is clear: the jump is often blurred by various reasons for strong glasses. This enables us to know the interrelations that hold among various energies related to the glass transition. From this, we are able to have coherent understanding of the physics behind the empirical relationship between fragility and specific-heat jump, which is described in Sec.~\ref{sec:discussion}.

%%%%%%%%%%%%%%%%%%%%%%%%%%%%%%%%%%%%%%%%%%%%%%%%
% \section{Methodologies}
\section{Calculation method of specific heat}  
\label{sec:Theory}

\subsection{Total-energy approach to specific heat}
\label{sec:formulationOfC}

\subsubsection{Specific heat of solids}
In standard textbooks of statistical mechanics, it is a common procedure to calculate specific heat of solids from phonon spectra on assumption of the harmonic approximation. Here, we have to start from general situations, because we need to treat a wide range of states of matter for which this approximation is no longer valid. Let us start a microscopic and dynamics theory in solids. Here, the microscopic theory means DFT, which is currently the most reliable theory to calculate ground-state properties of solids. The dynamic theory means MD simulations, which is needed for treating finite temperatures of course.

A solid is composed of $N$ atoms, whose positions are expressed as ${\bf R}_{j}$. 
The total energy $E_{\rm tot}$ of the solid is the sum of the potential energy $E_{P}$ and the kinetic energy $E_{K}$ of the atoms. In a microscopic timescale, both energies vary with time $t$,
\begin{equation}
E_{\rm tot}(t) = E_{P}(t) +  E_{K}(t),
\label{eq:total energy}
\end{equation}
where $E_{K} = (1/2) \sum_{j} M_{j} v_{j}^{2}$: $M_{j}$ and $v_{j}$ are the mass and velocity, respectively, of $j$th atom. The potential energy $E_{P}$ is what is called in DFT the ground-state energy $E_{\rm gs}$, which is a functional of atom positions $\{ {\bf R}_{j}(t) \}$ on the Born-Oppenheimer approximation. At equilibrium, an atom position at $t$ is given by the sum of the equilibrium position $\bar{\bf R}_{j}$ and the displacement ${\bf u}_{j}(t)$ from it, namely, ${\bf R}_{j}(t) = \bar{\bf R}_{j} + {\bf u}_{j}(t)$. Accordingly, $E_{P}(t) \equiv E_{\rm gs}( \{ {\bf R}_{j}(t) \} )$ can be expanded as
\begin{equation}
E_{P}(t) 
= E_{\rm gs}( \{ \bar{\bf R}_{j} \} ) +\frac{1}{2} \sum_{i,j}  {\bf u}_{i}(t) \cdot {\bf D}_{ij} \cdot {\bf u}_{j}(t),
\label{eq:potential_energy}
\end{equation}
where $\{ {\bf D}_{ij} \}$ is the force-constant matrix. 
The internal energy in the thermodynamics context, which is defined at equilibrium, is given by the time-average total energy, 
\begin{equation}
U =  \overline{ E_{\rm tot}( \{ {\bf R}_{j}(t) \} ) }.
\label{eq:UvsEtot}
\end{equation}
By using the expansion (\ref{eq:potential_energy}), we have
\begin{equation}
U(T, V, \{ \bar{\bf R}_{j} \})  = 
 E_{\rm gs}( \{ \bar{\bf R}_{j} \} ) + E_{\rm te}(V)
  +\frac{1}{2} \sum_{i,j} \overline{{\bf u}_{i}(t) \cdot {\bf D} \cdot {\bf u}_{j}(t)}
  + \frac{1}{2} \sum_{j} M_{j} \overline{v_{j}(t)^{2}},
\label{eq:internal_energy}
\end{equation}
with showing explicitly the arguments in $U$. The last two terms in Eq.~(\ref{eq:internal_energy}) have the same value because of the virial theorem. 
By converting the variables $u_{j}$ to the normal coordinates $q_{k}$, 
the sum of the last two terms turns to the phonon energy $E_{\rm ph} = \frac{1}{2} \sum_{k} \omega_{k}^{2} \overline{q_{k}^{2}}$ on the harmonic approximation. Here, $\omega_{k}$ is the frequency of $k$th phonon.
In Eq.~(\ref{eq:potential_energy}), $E_{\rm gs}( \{ \bar{\bf R}_{j} \} )$ is a function of $N$ coordinates $\{ \bar{\bf R}_{j} \} $. However, in Eq.~(\ref{eq:internal_energy}), extraction of six freedoms about the lattice parameters from the set $\{ \bar{\bf R}_{j} \}$ is understood, though the same notation is used. Henceforth, $\{ \bar{\bf R}_{j} \} $ means a set of $N-6$ coordinates of internal freedoms. For isotropic materials, the six parameters are presented by a single variable $V$. The thermoelastic (or thermal expansion) part $E_{\rm te}$ of the ground-state energy is singled out from the original $E_{\rm gs}$ in Eq.~(\ref{eq:potential_energy}). This part $E_{\rm te}$ has also absorbed the part of the phonon energy which has volume dependence. A systematic expansion of the total energy is described by Leibfried and Ludwig.\cite{Leibfried61} The remaining part $E_{\rm st}$ thus has no volume dependence: it is a function of only relative coordinates of a solid. We attach a name {\it structural energy} ($E_{\rm st}$) to the term $E_{\rm gs}( \{ \bar{\bf R}_{j} \} )$. In the glass literature, $E_{\rm st}$ is often called the configurational energy (Ref.~\onlinecite{Prigogine54}, p.~293); however the name of the structural energy presents directly its physical meaning and is preferable for the present study.
From Eq.~(\ref{eq:internal_energy}), we see that the internal energy $U$ is the sum of the structural, phonon, and thermal expansion energies as
\begin{equation}
U = E_{\rm st} + E_{\rm ph} + E_{\rm te}.
\label{eq:internal_energy2}
\end{equation}
When the volume is fixed, the structural energy $E_{\rm st}$ is obtained by
\begin{equation}
E_{\rm st} = \overline{E_{P}(t)} - \overline{E_{K}(t)} = 
  \overline{E_{\rm tot}(t)} - E_{\rm ph} = \overline{E_{\rm gs}(t)} - \frac{1}{2} E_{\rm ph}.
\label{eq:structural_energy}
\end{equation}
Given an internal energy $U$, the structural energy $E_{\rm st}$ is obtained by subtracting the phonon energy $E_{\rm ph}$ from $U$. 

Corresponding to the expression for the total energy, Eq.~(\ref{eq:internal_energy2}), the total specific heat, which is in turn isobaric specific heat $C_{p}$ in the thermodynamics context, is expressed as the sum of the structural,  phonon, and thermal expansion parts, as $C_{p} = C_{\rm st} + C_{\rm ph} + C_{\rm te}$. The first two components constitute the isochoric specific heat $C_{v} = C_{\rm st} + C_{\rm ph}$.
The contribution of the thermal expansion energy, $C_{\rm te}$, is known as (see for example Ref.~\onlinecite{Callen}, p. 189), 
\begin{equation}
  C_{\rm te} =\frac{T V}{\kappa} \alpha^{2}.
\label{eq:form-del_C_tv}
\end{equation}
In this study, two components $C_{\rm st}$ and $C_{\rm ph}$ are calculated by MD simulation with a constant $V$, while $C_{\rm te}$ is calculated from experimental data of $\alpha$ and $\kappa$.
At low temperatures, where the harmonic approximation well holds, there is no temperature dependence in the structural energy $E_{\rm st}$, and hence its contribution to specific heat $C_{\rm st}$ vanishes. The isochoric specific heat $C_{v}$ becomes the same as the phonon part, $C_{v}=C_{\rm ph}$: otherwise $C_{v}=C_{\rm st}+C_{\rm ph}$.

Calculation of the phonon part $C_{\rm ph}$ is a quite standard procedure. The phonon energy $E_{\rm ph}$ is calculated as
\begin{equation}
E_{\rm ph} = \sum_{k} \left( \bar{n}_{k} +\frac{1}{2} \right) \hbar \omega_{k}.
\label{eq:phonon-energy}
\end{equation}
Noticing that $T$ dependence appears through the Bose occupation number $\bar{n}_{k}= \left[ e^{\hbar \omega / k_{\rm B}T}-1 \right]^{-1} $ ($k_{\rm B}$ is Boltzmann's constant and $\hbar$ is Planck constant), we have for the phonon part $C_{\rm ph}$
\begin{equation}
C_{\rm ph}(T) = k_{\rm B} \int{  \left( \frac{\hbar \omega}{k_{\rm B}T} \right) ^{2} 
\frac{ e^{\hbar \omega / k_{\rm B}T} }{ \left( e^{\hbar \omega / k_{\rm B}T}-1 \right)^{2}} \ g(\omega) d\omega}
 \equiv
 k_{\rm B} \int{  f(\omega) g(\omega)} d\omega,
\label{eq:phonon-SH}
\end{equation}
where $g(\omega)$ is a phonon spectrum. Anharmonic effects have thus been taken into account at the level of the quasi-harmonic approximation.
A caveat is given that, even in FP-MD simulations, atom motions are treated in classical dynamics, meaning that the total specific heat calculated from the total energy contains the classical one $(3/2) R$ over all the $T$ range.
Hence, in the real implementation, first we remove the classical term $(3N/2) k_{\rm B}T$ from the potential energy as the first part of Eq.~(\ref{eq:structural_energy}), obtaining the structural part as $E_{\rm st} = \overline{E_{P}(t)} - \overline{E_{K}(t)}$. By calculating the phonon DOS $g(\omega)$ from time-dependent velocities, the phonon energy $E_{\rm ph}$ is obtained by Eq.~(\ref{eq:phonon-energy}). Finally, the total energy $U$ is obtained by the sum of $E_{\rm st}$ and $E_{\rm ph}$ so obtained. In this manner, low-temperature behavior of $C_{v}(T)$ is correctly reproduced.

When the harmonic approximation holds, the structural part $C_{\rm st}$ is hidden by the harmonic contribution. Change in the structural energy $E_{\rm st}$ comes to appear at high temperatures. In crystals, the maximum change in $E_{\rm st}$ occurs at the melting temperature $T_{m}$ as latent heat $H_{m}$. Since crystals have long-range order due to periodicity, the melting transition occurs at a single point of temperature $T_{m}$. For non-periodic systems, lack of long-range order alters the transition in a continuous manner, which renders the transition temperature broaden. A consequence of this continuous change is to bring about an excess specific heat $C_{\rm ex} = C_{p}^{(l)}-C_{p}^{(g)}$ in the temperature range $T_{g} < T < T_{m}$. Since $C_{p}^{(g)}$ is not a real observable in this range, $C_{p}^{(g)}$ in the formula is frequently substituted with the specific heat of crystal $C_{p}^{(c)}$. The excess enthalpy $H_{\rm ex}(T)$ as a function of $T$ is obtained by integration from $T_{g}$ to $T$. In particular, the excess enthalpy at $T_{m}$, 
\begin{equation}
H_{\rm ex} \equiv H_{\rm ex}(T_{m}) = \int_{T_{g}}^{T_{m}} C_{\rm ex}(T) dT = \overline{C_{\rm ex}} \ (T_{m}-T_{g}) ,
\label{eq:excess-enthalpy}
\end{equation}
can be interpreted to play a similar role as the latent heat $H_{m}$ does in solids: here $\overline{C_{\rm ex}}$ is the average excess specific heat in this region. The authors are, in the literature, aware of different views for the excess specific heat. However, it will be seen that holding the current view causes no contraction with any other part of the arguments in this paper. In the following, $H_{\rm ex}$ without arguments is referred to as $H_{\rm ex}(T_{m})$.

\subsubsection{Specific heat of liquids}
\paragraph{Phonon picture.}
There is no standard theory of specific heat for liquids. Only model descriptions are found in the literature. The most important fact towards constructing the standard theory may be that $C_{v}$ of simple liquids is close to the classical Dulong-Petit law, $C_{v} = 3 R$: throughout this paper, the gas constant $R$ is used in the unit of per mole of average atoms rather than per mole of molecules. 
This led researchers to consider that the phonon description is valid for liquids too.\cite{Wallace02}
The idea of phonons is of course fallacy for liquids for an obvious reason of lack of long-range order. Despite this, even for liquids, a vibration spectrum $g(\omega)$ exists in the sense that it is obtained by Fourier transform of atom velocities $v_{j}(t)$. By adapting the phonon formula (\ref{eq:phonon-SH}) to this vibration spectrum, the specific heat of liquid $C_{\rm ph}^{(l)}$ can be calculated. This method operationally works well, and the classical limit is reproduced by construction. 
In real situations, there is a small difference between this phonon contribution $C_{\rm ph}^{(l)}$ and the observed specific heat $C_{p}^{(l)}$. This difference is treated as the correction terms, such as anharmonic term and the boundary term by Wallace.\cite{Wallace98} However, the notions of these correction terms are not quite clear: do these terms exhaust all the possible corrections? There are a series of models in the line of an old thought by Frenkel \cite{Frenkel46}: diffusion of atoms in liquids is looked upon as successive jumps of atoms between many valleys of random potentials. Frenkel model has been developed by Granato \cite{Granato02} and Bolmatov {\it et al.},\cite{Bolmatov12} where the diffusion term plays a role. However, because of the model nature, it is difficult to make clear distinctions among many correction terms, and hence we are left uncertain whether the current calculation is overestimation or underestimation.

All the above methods are approaches to construct the observed quantity from elemental components. Such a bottom-up approach is correct if and only if all the components under consideration complete the whole sum and each component is independent to the others. For model calculations, there is no guarantee for this.
From the present view of DFT, the problem can be solved at one stroke. Only the total energy $E_{\rm tot}$ is the physically meaningful energy in DFT, and is the directly obtainable quantity.\cite{Parr-Yang89} 
This principle is valid also for liquids. The total internal energy $U$ is unambiguously obtained by Eq.~(\ref{eq:UvsEtot}) for any state of matter. Therefore, we can obtain the total specific heat $C_{p}$ (or $C_{v}$ under the constraint of constant volume) directly from calculation of the total energy, irrespective of whether it is a solid or is a liquid.
On the other hand, the perturbatic expression (\ref{eq:internal_energy}) is no longer valid for liquids. The time-averaged position $\overline{{\bf R}_{j}(t)}$ becomes indeterminate. In this case, the arguments in $U$ are $T$ and $V$ only, as is for ideal gases. For liquids, therefore, the decomposition of $U$ into Eq.~(\ref{eq:internal_energy2}) is physically not grounded. In our approach, $E_{\rm ph}$ is a virtual quantity that is {\em defined} by applying Eq.~(\ref{eq:phonon-energy}) to the vibrational spectra of a liquid, no matter whether the substance of phonon is real or not. The remaining part in $E_{\rm tot}$ defines $E_{\rm st}$, and hence is also a secondary quantity.

In the glass literature, researchers often study the phonon contribution $C_{\rm ph}^{(l)}$ to the specific heat of supercooloed liquids. In this case, the remaining part of specific heat $C_{\rm ex}^{(l)} = C_{p}^{(l)} - C_{\rm ph}^{(l)}$ is called the excess specific heat. The nature of the excess specific heat is long disputed in the glass physics. It is a common practice to look upon $C_{\rm ph}^{(l)}$ as the same as the specific heat of crystal $C_{\rm ph}^{(c)}$. In a simple thought, the difference in specific heat between the crystal and liquid comes from the freedom of atom configurations, which is called the configurational component of specific heat $C_{\rm cf}^{(l)}$. Accordingly, these two are equivalent: $C_{\rm ex}^{(l)} = C_{\rm cf}^{(l)}$. However, there are ever-ending arguments about the nature of $C_{\rm ex}^{(l)}$ as to whether $C_{\rm ex}^{(l)}$ has contribution from other sources than configuration, such as anharmonic effects and relaxation effects.\cite{Goldstein76,Gujrati80,Takahara95,Johari00,Starr01}
In the present view of DFT, the distinction is rather a matter of definition: only the total specific heat is a well-defined quantity. The same holds for entropy: only the total entropy $S$ is physically meaningful, while the configurational entropy $S_{c}$ is {\it defined} as the remaining part of $S$ after subtracting the vibrational part to entropy $S_{\rm ph}$.\cite{Han20}

\paragraph{Energy dissipation due to atom relaxation.}
Despite the similarity in the phonon treatment between solids and liquids, there is an essential difference. Phonons are not eigenstates in liquids. Each vibration mode $k$ is subjected to energy dissipation processes. It has a finite lifetime $\tau_{k}$: phonons in solids have also finite lifetimes; however, this does not invalidate the notion of phonons for solids, because the lifetimes are so long. The energy dissipation processes in liquids can be looked upon as phonon-phonon conversion among different normal modes. 
The phonon conversion is an irreversible process, which accompanies energy dissipation. This energy dissipation is reflected in viscose flow of fluids. Since in the transition region viscosity grows exponentially, this effect cannot be ignored when glass transition is studied.
% This region is the energy converting process between the potential and kinetic energy. The relaxation time of this conversion is longer than the thermal relaxation time, the process becomes irreversible one.
%
In solids, the effects of finite lifetime of phonons on the optical susceptibility $\chi$ appear as the shift in phonon frequencies and broadening.\cite{Cowley64,Reissland-phonon} The energy dissipation means $\omega$ dependence of $\chi(\omega)$, which yields the imaginal part $\chi(t)$. 
In the spirit of the linear response theory, specific heat $C$ is regarded as the susceptibility for the small perturbation of temperature. Accordingly, the specific heat also exhibits $\omega$ dependence. This was for the first demonstrated by Birge and Nagel.\cite{Birge85} Formal theory of $\omega$-dependent specific heat $C(\omega)$ has been established.\cite{Nielsen96,Hentschel08} However, the linear-response theory itself does not provide practical calculation methods for $C(\omega)$. Information about the lifetime $\tau_{k}$ of each vibration mode $k$ is needed. For solids, calculation of the lifetime has been established on the basis of perturbation theory. \cite{Maradudin62} 
For liquids, the effect of lifetime is so large that the perturbatic treatment is inappropriate.
Direct application of the fluctuation-dissipation relation can provide a calculation method for susceptibility without knowing $\{ \tau_{k} \}$.\cite{Shinoda05} But, in the present implementation, it is unclear how the contribution of the structural energy is coped in the formula.

The above difficulties can be solved, at one stroke, by adiabatic MD simulations based on the total-energy approach. We do not need to know lifetimes $\{ \tau_{k} \}$ but only their consequences on the equilibrium states are sufficient for calculating specific heat. The maximum-entropy principle guarantees that entropy $S$ is uniquely determined by a given $U$ when equilibrium is established under given constraints, from which $T$ is uniquely determined by $T=\partial U/\partial S$. The total energy $E_{\rm tot}$ is conserved in adiabatic MD simulation. Accordingly, an infinite number of energy-dissipation processes finally lead the system to the unique state of equilibrium with $T_{\rm eq}$, which may be different from the initial temperature. All the effects of energy dissipation processes are included in the relation between the total internal energy $U$ and the final temperature $T_{\rm eq}$ in equilibrium, whatever energy dissipation processes are complicated.
In this regards, use of adiabatic MD simulations is essential. When the temperature is controlled by introducing a heat bath, such as Nos\'{e}-Hoover thermostat, energy dissipations to the artificial heat bath destroy the intrinsic relationship between $U$ and $T$, which is a property of a given material. In this respect, the present method is easy, because neither $T$ nor $V$ is needed to control.

\subsection{Implementation of MD simulations to glass transition}
\label{sec:MDsimulation}
As mentioned in Introduction, serious obstacles of study on glass transition by MD simulations are the timescale and hysteresis problems. Both problems are related each other. Hysteresis is caused by retardation of atom rearrangement (atom relaxation) against the change in an external influence, such as temperature. The timescale of glass transition is a microscopic time longer than 1 s, and hence it is a hopeless task to perform a faithful simulation for the whole process of transition even by model calculations.
This difficulty can be evaded by simulating the whole process of the transition by a series of successive MD runs with changing the total energy from run to run. In an adiabatic MD run, say $k$th run, given the internal energy $U^{(k)}$ with the initial positions of atoms $\{ {\bf R}_{j}^{(k)}(0) \}$, an adiabatic MD simulation is performed until an equilibrium is reached with the equilibrium temperature $T^{(k)}$.

At this stage, it may be needed to explain what the equilibrium means for the glass transition, because the glass state is usually considered as a nonequilibrium state. In practice, we have no difficulty in obtaining the so-called equilibrium temperature $T$ in MD simulations. In each of the present adiabatic runs, we always observed that the system finally reached a steady state in the sense that the time average of kinetic energy $\langle E_{K}(t) \rangle$ in a short period is constant, from which the equilibrium temperature $T$ is determined. Thus, the temperature so obtained is a well-defined quantity and is a thermodynamic state variable. For the liquid state, the internal energy $U$ is uniquely determined by $T$, and accordingly $U=U(T)$ becomes a thermodynamic relation, as should be.
On the other hand, for the solid state, different values were obtained for $U$ even at the same $T$: the value of $U$ depends on the past history. This entails that, if $T$ were only a state variable, $U=U(T)$ would not be a thermodynamic relation. 
However, if the equilibrium state in the above sense is established when the liquid freezes in, it is obvious that $U$ is uniquely determined by a set of the equilibrium positions $\{ \bar{\bf R}_{j} \}$, irrespective of the past history: certainly, the current values $\{ \bar{\bf R}_{j} \}$ are the consequence of the past history; but 
the current value of $U$ is uniquely determined by the current values of $\{ \bar{\bf R}_{j} \}$ solely once the equilibrium positions $\{ \bar{\bf R}_{j} \}$ were fixed. Therefore, $U = U(T, \{ \bar{\bf R}_{j} \})$ must be a thermodynamic relation. This infers that equilibrium positions $\{ \bar{\bf R}_{j} \}$ have a role  as state variable for glasses (more generally, solids). In fact, this is proven from the thermodynamic grounds.\cite{Shirai20-GlassState,StateVariable} 
It is an observation of experimentalists that, if experiments of the glass transition are carefully carried out, it is possible to suppress irreversibility at minimum in order to get good reproducibility in measuring thermodynamic properties.\cite{Goldstein76,Goldstein08,Johari10}

For the solid state, the equilibrated positions $\{ \bar{\bf R}_{j}^{(k)} \}$ of $k$th run are obtained by time average after equilibrium is reached, and the so-obtained positions are passed to the initial positions of the next $(k+1)$th run. For the liquid state, the last positions $\{ {\bf R}_{j}^{(k)}(t_{\rm SM}) \}$ of $k$th run are passed to the next run ($t_{\rm SM}$: the total simulation time).
At the same time, the internal energy $U^{(k+1)}$ is slightly changed: this can be done by changing the initial kinetic energy $E_{K}^{(k+1)}$, as usual. The amount of change $\Delta E_{K}^{(k+1)}$ gives a rough estimate for the temperature change $\Delta T^{(k+1)}$ in the successive runs. Positive $\Delta E_{K}^{(k+1)}$ corresponds to heating, and negative one corresponds to cooling. It is of course meaningless to speak of the rate of change in $T$, because there is no idea of time span between successive runs. However, by controlling the temperature interval $\Delta T^{(k+1)}$, we can mimic the rate dependence of cooling/heating.
The present procedure of successive adiabatic MD runs is what is done in experiments of the adiabatic calorimetry and the drop calorimetry.\cite{Davies53a,Davies53}

%%%%%%%%%%%%%%%%%%%%%%%%%%%%%%%%%%%%%%%%%%
\section{Application to glycerol}
\label{sec:Application}
%%%%%%%%%%%%%%%%%%%%%%%%%%%%%%%%%%%%%%%%%

Let us to apply the foregoing method to a molecular glass of glycerol. In Fig.~(\ref{tab:Properties-Glycerol}), experimental data measured by Gibson and Giauque is shown.\cite{Gibson23} The glass-transition temperature is 185 K. The jump in specific-heat is identified as $\Delta C_{p} = 0.70 R$.
The properties of glycerol which are relevant to this study are summarized in Table \ref{tab:Properties-Glycerol}.

\subsection{Calculation conditions}
\label{sec:condition}

\begin{table}
\begin{tabular}{l  r l l}
  \hline \hline
 Property & Value  & & Ref. \\
  \hline
  Chemical formula & \multicolumn{2}{c}{ ${\rm C_{3}H_{8}O_{3}}$}  & \\
  Density, $\rho$ & 0.76/0.75  &  cm$^{3}$/g  & $^{a}$ \\ % Moynihan81 
%  Molar volume, $v$ & 73.0/65.4  &  cm$^{3}$/mol  & a, b \\ % 
  Melting temperature, $T_{m}$ &  291.2 & K & $^{b}$ \\ % CRC
  Latent heat of fusion, $H_{m}$  &  18.3 & kJ/mol & $^{b}$ \\  % CRC, ? 12.26$^{(b)}$ 
  Glass-transition temperature, $T_{g}$ &  185 & K & $^{c}$ \\ % Rao02
  % Specific-heat jump, $\Delta C_{p}$ &  0.212 & cal/g.deg \\ % Rao02
  % 0.86 J/g.deg = 79.12 J/mol.deg = 0.86 J/atom.deg = 0.103 k_B/atom
  Specific heat, $C_{p}$ & 1.91/1.05 & J/g.deg & $^{d}$ \\ % Kauzmann 0.458/0.25 cal/g.deg
  Thermal expansion, $\alpha$ &  4.83/2.4  & $\times 10^{-4}$/deg & $^{d}$ \\ % Kauzmann
  Compressibility, $\kappa$ &  0.18/0.10 & /GPa & $^{e}$ \\ % Christensen94
%  Configuration entropy, $S_{c}(T_{g})$ &  26.7 & J/K.mol & b \\ % Rao02
% \UTF{2206}Hg-c = 7.52 & kJ/mol
    \hline  
\end{tabular}
\caption{Thermodynamic properties of glycerol. Data are take from: (a) Ref.~\onlinecite{Moynihan81}, (b) Ref.~\onlinecite{CRC92}, (c) Ref.~\onlinecite{Rao02}, (d) Ref.~\onlinecite{Kauzmann48}, (e) Ref.~\onlinecite{Christensen94}. Two values appearing in an entry correspond to those of liquid/glass states. }
\label{tab:Properties-Glycerol}
% (b) Gibson23, Parks34
\end{table}

\begin{figure}[htbp]
    \includegraphics[width=130mm, bb=0 0 420 300]{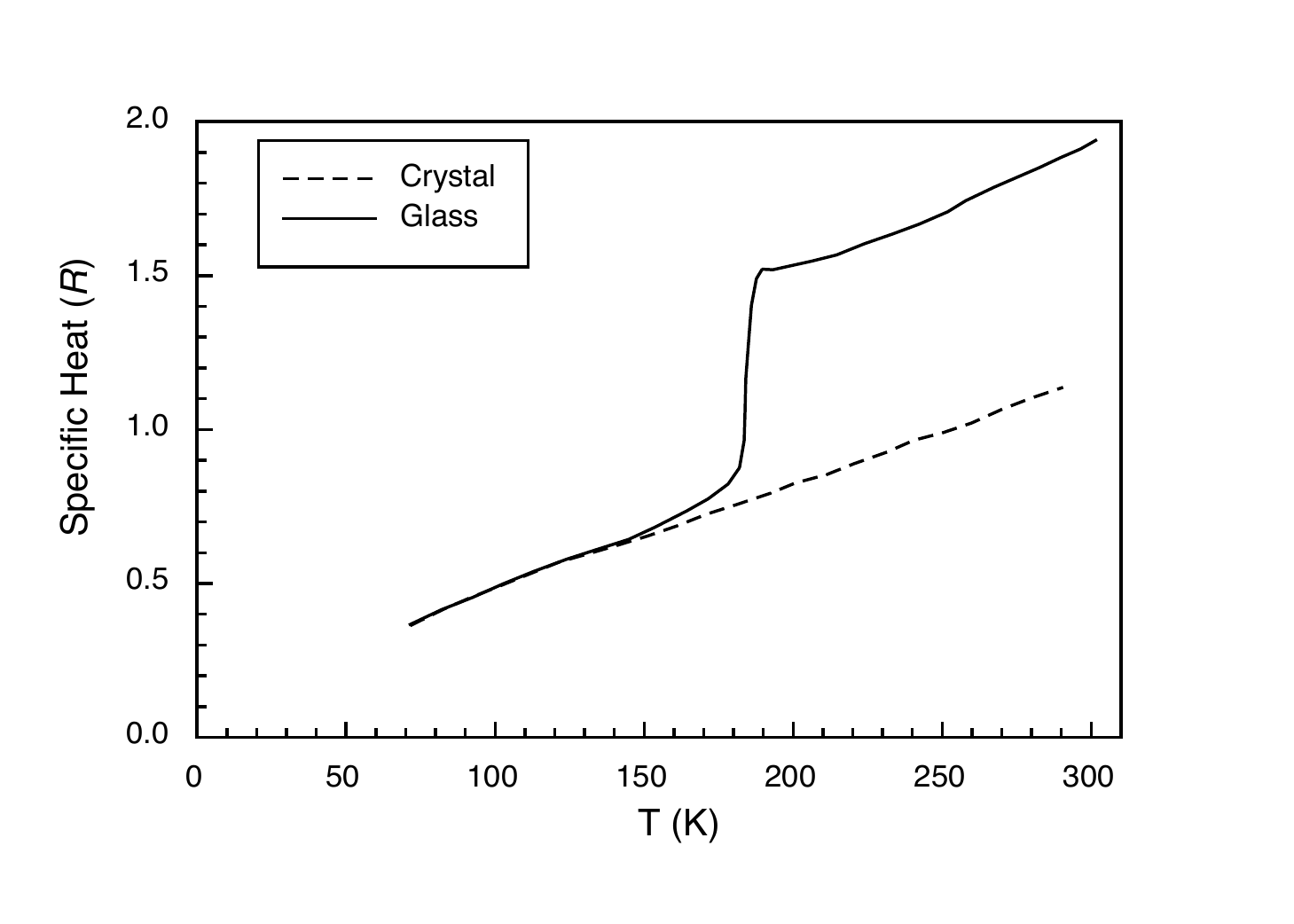} 
  \caption{Isobaric specific heat $C_{p}$ of glycerol per mole of atoms. The figure are redrawn from the data by Gibson and Giauque \cite{Gibson23} with rescaling in the unit of $R$. } \label{fig:tot-sh-T}
\end{figure}
  % Specific-heat jump, $\Delta C_{p}$ &  0.212 & cal/g.deg = 81.5 J/mol.deg = 5.82 J/atom.deg = 0.70 kB
  % Rao02

The used code for FP-MD simulations is Phase/0.\cite{PHASE} It is a pseudopotential method: oxygen atom is treated by a ultrasolf potential, while silicon by a norm-conserved potential. The cutoff energy for the expansion of planewaves is 30 Ry. One-point ($R$ point) $k$ mesh is used. The time step in MD simulations is $40$ atomic time. The total simulation time $t_{\rm SM}$ varies from 4 to 12 ps: when the presence/absence of diffusion obvious, 4 ps is enough, while when it is marginal, long $t_{\rm MD}$ is used.
Volume $V$ is fixed throughout a series of MD simulations.

As usual, the temperature $T$ in MD simulations is obtained by time and particle average of the kinetic energy of atoms $\langle \overline{E_{K}(t)} \rangle$. The initial MD steps were discarded until a constant $T$ is achieved. In most cases, the relaxation of temperature was fast, namely, an order of 0.1 ps. In the worst case, a few ps was required, which happened in the transition region.
Reaching equilibrium is judged by examining $t$ dependence of the averaged displacements $\overline{x_{j}(t)^{2} }$. When $\overline{x_{j}(t)^{2} }$ is constant with respect to $t$, it is an equilibrium state of the solid phase. When linearity between $\overline{x_{j}(t)^{2} }$ and $t$ is observed over the entire simulation time $t_{\rm SM}$, it is an equilibrium state of the liquid phase. The slope of the linear relationship gives diffusion coefficient $D$.
If $\overline{x_{j}(t)^{2} }$ varies with $t$ but not yet establish the linear relation, the system is in the transition region. Usually, the transition states were observed only within a narrow $T$ range, so that $T_{g}$ is easily identified. 

The glass state of glycerol is obtained by heating the crystalline phase. The structural parameters are taken from Ref.~\onlinecite{Koningsveld68}. After melting at sufficiently high temperatures, this liquid is cooled down. Details of dynamics of structural changes are given elsewhere.

\subsection{Heating/cooling simulation}
\label{sec:processes}

\begin{figure}[htbp]
    \centering
    \includegraphics[width=120mm, bb=0 0 430 500]{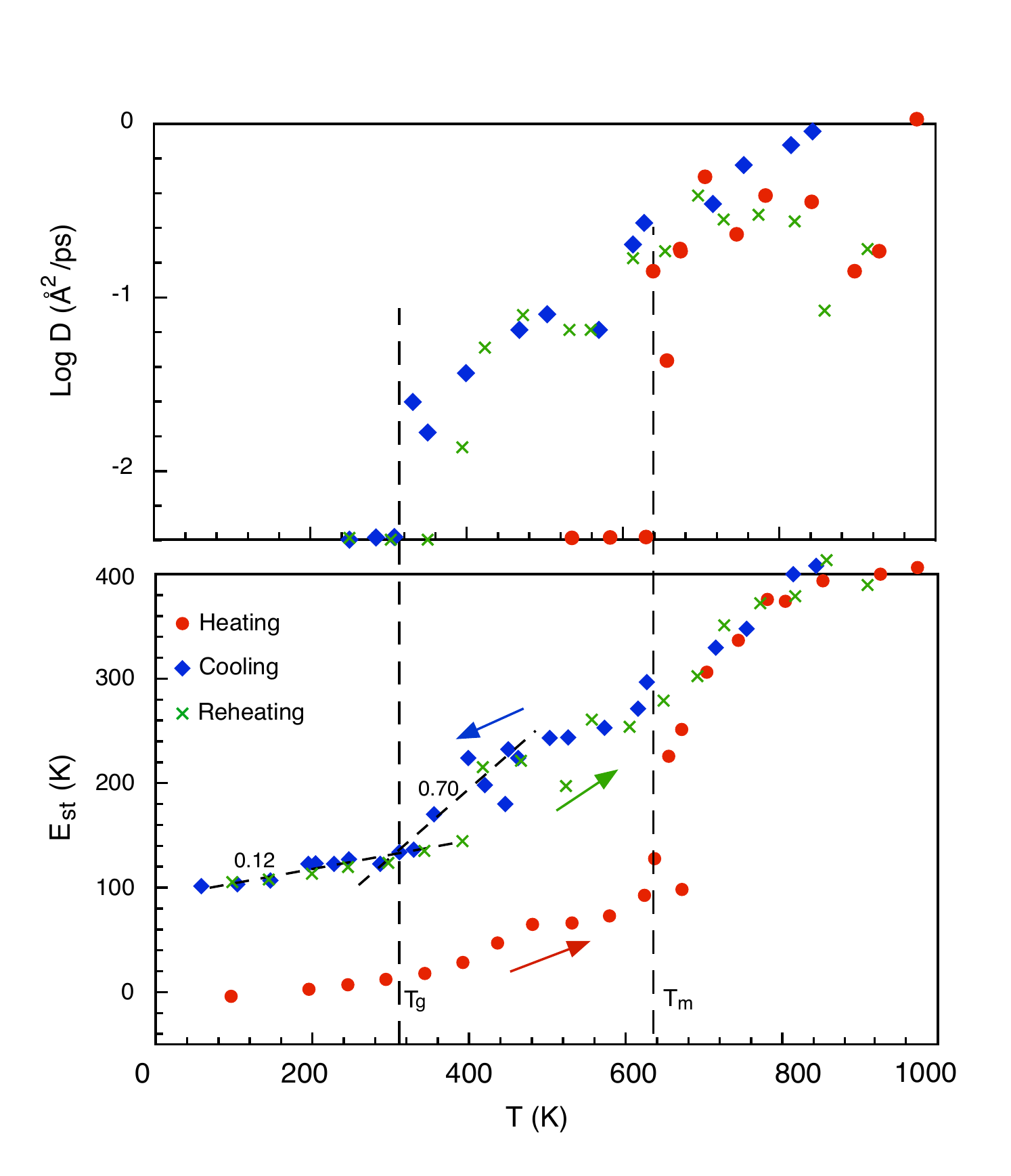} 
  \caption{Diffusion coefficient $D$ and structural energy $E_{\rm st}$ as a function of $T$ in heating, cooling, and reheating processes.
Data points of $\log D =-2.4$ actually indicate $D=0$ within the accuracy of the present simulations. These are plotted only because it makes it clear where glass transition occurs. 
% The energy origin is taken to the ground-state energy of crystal glycerol.
  } \label{fig:D-Est}
\end{figure}

Our simulation started from crystalline glycerol at low temperatures; it was heated up to melt and then the melt is cooled down to obtain glass state. At last, the glass samples were reheated. We study the glass transition mainly in the cooling process, because it renders the influence of the energy barrier minimum: generally the energy barrier is smaller in cooling process than in heating process. The last step of reheating was added merely to check the hysteresis.
Figure \ref{fig:D-Est} shows variations of diffusion coefficient $D$ and the structural energy $E_{\rm st}$ as $T$ is varied. The temperature $T$ in this figure and in all the subsequent figures indicates the average temperature after reaching thermal equilibrium. In all the runs, $T$ is well converged to a unique value, in most cases within a few tenths of ps. (We do not claim that, in the glass transition, the entire equilibrium is reached in such short times: every thermodynamic variable has its own relaxation time; there are fast and slow variables; phonon motions are fast variables and therefore the above thermal equilibrium merely means that the phonon subsystem quickly reaches equilibrium. The atom relaxation requires much longer times).\cite{Shirai20-GlassState} In the figure, $E_{\rm st}$ is presented in a unit K/atom, which makes it easier to see specific heat in $R$ units. 

Let us first examine the heating process. Although there is large fluctuation in $D$, melting is clearly seen at about $T=635$ K. The calculated value is quite largely deviated from the experimental one (291 K). In our experience on MD simulation on, for example, silica, overestimations of this order of magnitude are commonly observed in simulations of melting by FP methods. One reason is that periodic boundary conditions removes all the surface effects, which are have an important role on inhomogeneous nucleation. Another reason is the effect of spurious reflections for atom movements, which are caused by the artificial periodic-boundary condition. We consider that the latter effect is more serious when the size of supercell is such a small one as the present case. Introduction of disorder is expected to have a similar effect as the surface effect with respect to creation of nuclei of the solid phase. This effect acts on irrespective of heating or cooling. The simulation result, as shown later, is different from this expectation. 
Therefore, the overestimation of $T_{m}$ is brought about by the spurious reflection due to use of small supercells. Further analysis of this effect is given elsewhere.\cite{Shirai22-Silica} By considering feasible size of systems for the present FP-MD technique to treat, this is an unavoidable error, and therefore we leave this problem untouched. The readers should read the following description with keeping this overestimation in mind.

At low temperatures, the structural energy $E_{\rm st}$ is almost independent of $T$, indicating that the potential shape is nearly a harmonic form. As $T$ increases, $E_{\rm st}$ gradually increases, and then shows a jump at $T=635$ K. The agreement of this temperature with the value $T_{m}$ obtained by the diffusion coefficient confirms this jump as the latent heat of melting. The calculated value $H_{m}$ is in a range from 100 to 150 K/atom (from 11.5 to 17.4 kJ/mol, owing to a finite width around $T_{m}$. In this range, the system is dynamically unstable.\cite{Shirai22-Silica} In spite of this, the calculated value $H_{m}$ is not bad compared with the experimental value (18.3 kJ/mol).
% 1680 K/mol = 0.1448 eV/mol = 13.94 kJ/mol

Let us next examine the cooling process from the liquid.
The diffusion coefficient $D$ decreases with decreasing $T$. At $T=310$ K, it vanishes within the present accuracy. Correspondingly, the slope of $E_{\rm st}$ with respect to $T$ changes at this temperature. Hence, this temperature is identified as the glass-transition temperature, $T_{g} = 310$ K. This value is again higher than the experimental value of 190 K. But by considering the above-mentioned errors due to the small size of supercell, the calculated value is reasonable.
Large hysteresis appears in the transition region. When the sequence of $\{ \Delta T^{(k)} \}$ was altered, different results were obtained even at the same temperature. It is necessary to keep the temperature step $\Delta T^{(k)}$ as small as possible and to change $T^{(k)}$ in a monotonous sequence: otherwise the system is easily trapped in a local minimum of potential. The data shown in Fig.~\ref{fig:D-Est} are obtained with paying the above cares. The data points are still not smooth enough to make numerical derivative meaningful. 
Hence, we content ourselves with only evaluating $\Delta C_{\rm st}$ by linear fitting of data $E_{\rm st}$ above and below $T_{g}$. These linear fittings are indicated by dashed lines in Fig.~\ref{fig:D-Est}. 
From this, we obtain for the contribution of the structural energy to the specific-heat jump,
\begin{equation}
\Delta C_{\rm st} = 0.58 \ R.
\label{eq:del_C_st}
\end{equation}
This value is comparable with the observed jump $0.70 \ R$. This indicates that the jump $\Delta C_{p}$ is almost determined by the contribution of the structural energy.
The width $\Delta T_{g}$ of glass transition cannot be identified in the present accuracy of simulation.

The enthalpy difference $\Delta H_{gc}$ between the glass and crystal is obtained from extrapolation of $E_{\rm st}$ to $T=0$, if the contribution of thermal expansion is ignored. From Fig.~\ref{fig:D-Est}, we obtain $\Delta H_{gc}=11.5$ kJ/mol (8.6 meV/atom).
% 0.00862 eV/atom = 0.120 eV/mol
For the experimental data, $\Delta H_{\rm gc}$ is obtained by integrating the specific heat and the latent heat. By using Gibson and Giauque\cite{Gibson23} shown in Fig.~\ref{fig:tot-sh-T}, we obtain 9.5 kJ/mol. The agreement in $\Delta H_{gc}$ between calculation and experiment is good. 

As the last step of simulations, reheating of the glass was added, as indicated by cross marks in Fig.~\ref{fig:D-Est}. The melting of the glass begins at about $T=420$ K. This value is still much higher than $T_{g}$ that is obtained in the cooling process. This result is the evidence that the surface effect is not the main cause of the large overestimation in $T_{m}$ and $T_{g}$, as described above.
Some reader might suspect that the difference in $T_{g}$ between the cooling and reheating processes indicate the hysteresis that is commonly observed in experiments. However, the hysteresis in the present simulation is a different kind from that of experiments. First, although different $C$-$T$ curves are observed in experiments between cooling and heating processes, the higher bound of glass transition $T_{g2}$ is almost the same in a reasonable range of the conditions of measurement.\cite{Moynihan76,Hodge94} Second, the experimentally observed hysteresis is caused by inhomogeneity in spatial scales much larger than the size of the present supercells.

\subsection{Specific heat}
\label{sec:specific-heat}

\begin{figure}[htbp]
    \centering
    \includegraphics[width=120mm, bb=0 0 380 260]{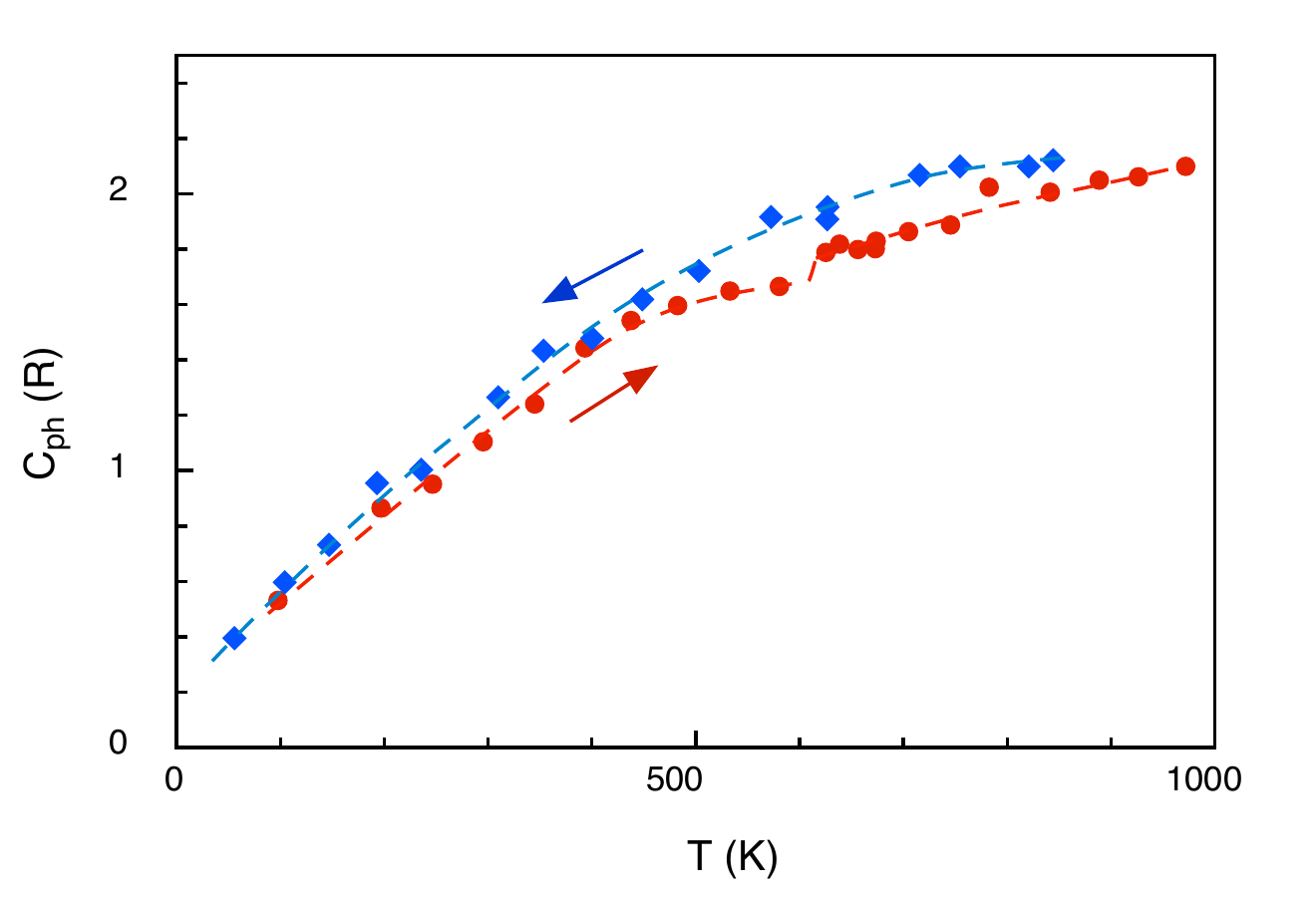} 
  \caption{Phonon contribution to specific heat $C_{\rm ph}$. Red circles indicate the heating process from the crystal. Blue rhombs indicate the cooling process from the liquid. Interpolation curves are drawn by dashed lines.} \label{fig:sh_ph-T}
\end{figure}

\paragraph{Phonon contribution.}
The phonon contribution to specific heat, $C_{\rm ph}$, was calculated by Eq.~(\ref{eq:phonon-SH}) and the result is shown in Fig.~\ref{fig:sh_ph-T}.
In the heating process, a small discontinuity is observed in the $C_{\rm ph}$-$T$ curve at about $T=600$ K, which shows the melting of crystal. In contrast, no clear discontinuity is observed in the cooling process. The jump $\Delta C_{\rm ph}$ is blurred by the fluctuation in the calculated $C_{\rm ph}$. Hence, the jump $\Delta C_{\rm ph}$ is estimated to be less than $0.02 R$.

In the temperature range $T_{g} < T < T_{m}$, the calculated $C_{\rm ph}^{(g)}$ is larger than $C_{\rm ph}^{(c)}$, as is observed in experiments. However, in the range of the normal-liquid state, $T>T_{m}=635$ K, the calculated values of $C_{\rm ph}^{(l)}$ do not match between the heating and cooling processes. Even at $T>T_{m}$, our simulations might have influence of the small size and/or the limited time of simulations.

\begin{figure}[htbp]
    \centering
    \includegraphics[width=140mm, bb=0 0 450 480]{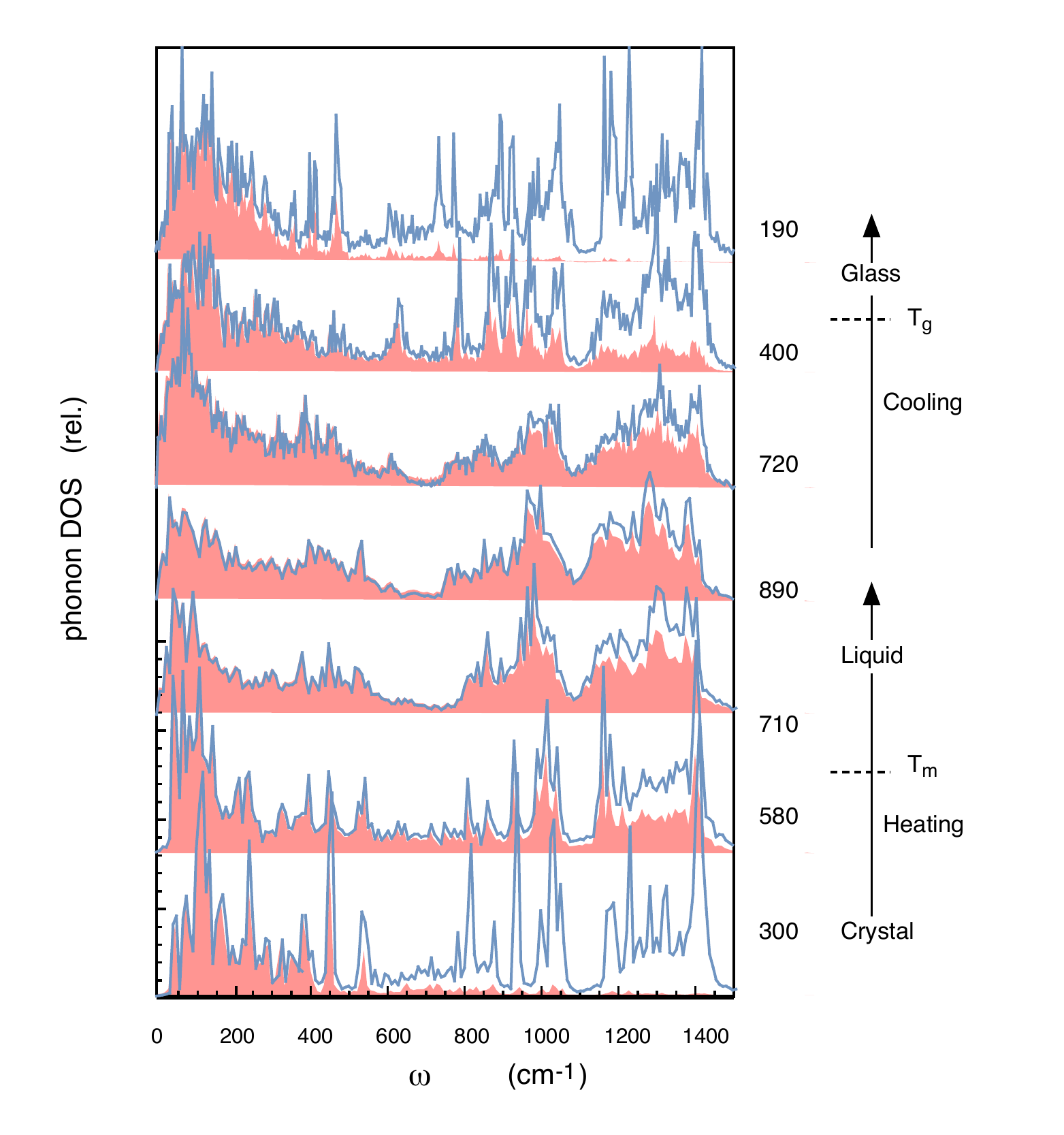} 
  \caption{Phonon spectra and its thermal occupancy of crystal and glass in heating from the crystal and in cooling from the liquid. Aside from the shown range, the hydrogen-related bands appear in the range $\omega>3000\ {\rm cm}^{-1}$. The read areas indicate the fraction of phonons that are thermally activated.} \label{fig:ph-spectra}
\end{figure}

In Fig.~\ref{fig:ph-spectra}, the selected phonon spectra $g(\omega)$, from which the specific heat $C_{\rm ph}$ was calculated, are shown. 
The degree how largely the mode of frequency $\omega$ is thermally activated can be represented by the factor $f(\omega)$ in Eq.~(\ref{eq:phonon-SH}): it varies from 0 to 1 as $T$ increases from 0 to infinite. In the figure, the red area indicates this degree of thermal activation. There are high-frequency bands $\omega>3000\ {\rm cm}^{-1}$, which are due to hydrogen stretching motions. These phonons are thermally inactivate and hence omitted from the figure.
In the heating process, the sharp feature at $\omega>1100 \ {\rm cm}^{-1}$ is broadened, while the integrated DOS in the range $700 < \omega < 1100 \ {\rm cm}^{-1}$ increases. This change is the primal cause for the small discontinuity in $C_{\rm ph}$ at melting of crystal. On the other hand, there is no abrupt change in phonon spectra when the glass is formed in the cooling process. 

\paragraph{Total specific heat.}
The contribution of thermal expansion $\Delta C_{\rm te}$ is calculated by Eq.~(\ref{eq:form-del_C_tv}).
The data needed to evaluate this formula are listed in Table I, from which $\Delta C_{\rm te} = 0.09R$ is obtained. The contribution of thermal expansion to $\Delta C_{p}$ is negligible.
% 0.093R
% \begin{equation}
%  \Delta C_{\rm te} = 0.093 \ k_{\rm B}\label{eq:del_C_tv}
% \end{equation}
% \cite{Callen85}
%
From summing all the three components, we obtain
\begin{equation}
\Delta C_{p} = 0.67 R.
\label{eq:del_C_tot}
\end{equation}
The experimental value for the specific-heat jump is $\Delta C_{p}^{\rm exp} = 0.70R$.\cite{Gibson23,Rao02} Accordingly, there is a good agreement with the experimental value. The most part of $\Delta C_{p}$ comes from the contribution from the structural part $\Delta C_{\rm st}$: 87 \% of the total $\Delta C_{p}$.

The specific heat of the supercooled liquid state is noted. As seen in Fig.~\ref{fig:tot-sh-T}, in a range $200 <T < 300$ K, the experimental $C_{p}^{(l)}$ increases linearly in $T$: by $0.4 R$ per 100 K. In calculation, in the range $T_{g} < T < T_{m}$, there is no appreciable increase in the slope of $E_{\rm st}$, as seen in Fig.~\ref{fig:D-Est}, so that $C_{\rm st}^{(l)}$ is negligible. On the other hand, the phonon contribution $C_{\rm ph}^{(l)}$ in the corresponding range increases by $0.33 R$ per 100K. The contribution of thermal expansion $C_{\rm te}^{(l)}$ is $0.16 R$ per 100K. In gross, an increase of $0.49 R$ per 100K is obtained for the increase in $C_{p}^{(l)}$. This again overestimates a little the experimental value. However, it is certain that the increase of $C_{p}^{(l)}$ with increasing $T$ is dominated by the phonon contribution.

%%%%%%%%%%%%%%%%%%%%%%%%%%%%%%%%%%%%%%%%%
\subsection{Discussion}
\label{sec:discussion}
%%%%%%%%%%%%%%%%%%%%%%%%%%%%%%%%%%%%%%%%%

\subsubsection{Thermodynamics consideration}
\label{sec:thermodynamics}
\paragraph{Factors determining the specific-heat jump.}
In this section, the significance of the preceding conclusion that the specific jump $\Delta C_{p}$ is dominated by the structural energy is discussed. Several models for explaining $\Delta C_{p}$ are seen in the literature: see the review of Nemilov (Chap.~II of Ref.~\onlinecite{Nemilov-VitreousState}). There, the specific-heat jumps for several materials are explained on the basis of structural models. Here, we discuss it from more general point of view.
There is an old interpretation for the jump, \cite{Thomas31} which is even now held in some researchers' minds. If a glass has the transition temperature $T_{g}$ lower than the Debye temperature $\theta_{D}$, the specific heat of the glass at $T_{g}$ is less than the classical limit $3R$. Since glasses are solids, atom positions are constrained by chemical bonds. When the glass melts, these constraints are removed, and softening of the phonon spectrum occurs. This causes a sudden increase in the thermal occupancy at high-frequency modes, which were not occupied before the transition, resulting in an increase in $C_{p}^{(l)}$ towards $3R$. If $T_{g} \gg \theta_{D}$, since thermally occupation is deeply saturated, a slight change in $T$ does not cause any change in $C_{p}$. The fact of general tendency that the specific-heat jump $\Delta C_{p}$ is large for fragile glasses ($T_{g} \ll \theta_{D}$ in most cases) while is small for strong glasses ($T_{g} \approx \theta_{D}$) is in favor of this interpretation.
However, the present result shows that this conventional interpretation is not supported. Certainly softening of the phonon spectrum occurs when the glass melts, but the change is only gradual over the range of $\Delta T_{g}$, and therefore the effect on specific heat is insignificant. 
Essentially the same conclusion was already deduced by experiment\cite{Smith17} and MD simulation\cite{Han20} for metallic glasses, although the different term entropy is used in place of energy. The description in terms of energy has a merit to explain the issue of fragility, which becomes soon clear.

The contribution from thermal expansion $C_{\rm te}$ is also insignificant in determining $\Delta C_{p}$. Therefore, we obtain an approximate relation $\Delta C_{p} \approx \Delta C_{v}$. The jump $\Delta C_{p}$ is, in a good approximation, determined by the change in the structural energy $E_{\rm st}$. 
The relation of $\Delta C_{p} \approx \Delta C_{v}$ for the glass transition is at variance with the phase transitions of the second-order type, where there is no appreciable variation in $C_{v}$ (see Ref.~\onlinecite{Pippard}, Chap.~9). In fact, the activation energy $Q_{a}$ is associated with any glass transition, which implies a discontinuous change in the structure. In this sense, the glass transition can be classified as the first-order transition. There are, however, differences from normal melting/crystallization at points of the sizable magnitude of energy barrier, which makes the transition irreversible and gives rise to hysteresis, and the randomness of potential, which breaks up the latent heat to spread over a finite range of temperatures, creating the excess specific heat $C_{\rm ex}$. Classification of phase transitions must be made with care, as noted by Pippard.\cite{Pippard,Pippard85} After all, we will find that the normal classification according to Ehrenfest is a theory based on reversible transitions, in which the above two properties are excluded. 
% Hence, a long-debated issue whether the glass transition is classified as the first-order transition or not is to some extent a problem fo definition.
We do not go further to this issue however, because phase transition is out of the present scope.
% This conclusion shares a common idea with the random first-order transition \cite{Lubchenko07, Mezard12} at the point of the presence of energy barrier. 

\begin{figure}[htbp]
   \centering
    \includegraphics[width=100mm, bb=0 0 600 500]{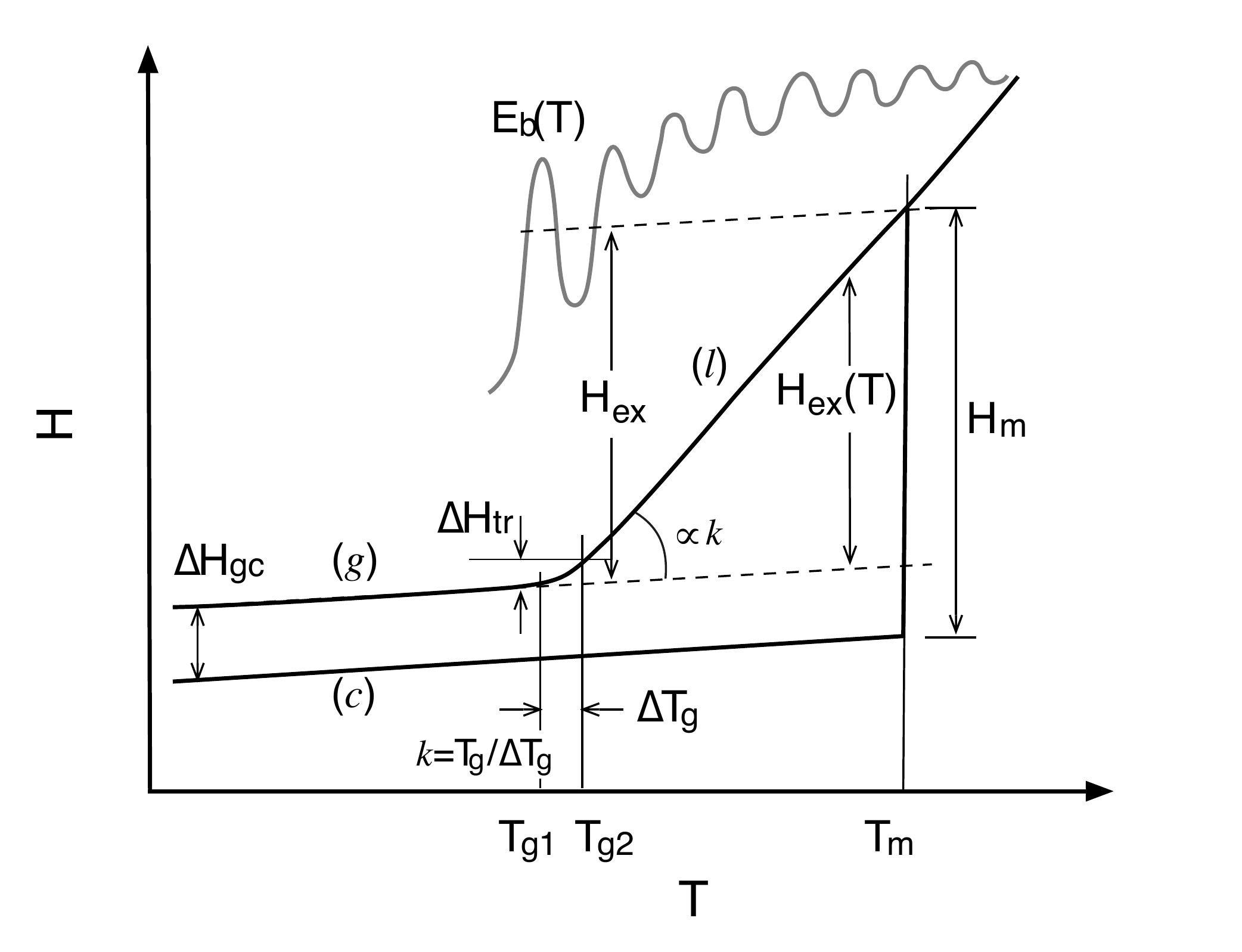} 
  \caption{
  Energy scheme of the glass transition. 
  Shown energies are the latent heat $H_{m}$, the enthalpy difference $\Delta H_{gc}$ between the glass and crystal, and the transition enthalpy $\Delta H_{\rm tr}$ at $T_{g}$, as well as the $T$ dependent excess enthalpy $H_{\rm ex}(T)$. The $T$ dependence of energy barrier $E_{b}$ is also indicated by a gray curve. The energy scale of $E_{b}$ is not adjusted with that of enthalpy. Its oscillatory behavior as a function of $T$ is not real but the illustration is intended to show the random feature of energy barriers.
  } \label{fig:HeatCycleGT}
\end{figure}

\paragraph{Energy scheme of glass substance.}
Let us discuss relationships that hold among various energies related to the glass transition. These energies are shown in the energy scheme in Fig.~\ref{fig:HeatCycleGT}. For crystals, the size of latent heat for melting is close to $RT_{m}$, 
\begin{equation}
H_{m} \approx R T_{m}.
\label{eq:Hm-RTm}
\end{equation}
This can be checked by numerous data on thermodynamics (Ref.~\onlinecite{CRC92}, p.~6-146 and 12-206). Because melting/crystallization is a reversible process, the energy barrier of nucleation when cooling from liquid is normally negligible.
In the glass transition, the excess enthalpy $H_{\rm ex}$ plays a similar role as the latent heat $H_{m}$ in crystal, as is described in Sec.~\ref{sec:formulationOfC}. The glass transition temperature $T_{g}$ is empirically known as about two third of the melting temperature $T_{m}$.\cite{Berthier11} Hence, Eq.~(\ref{eq:excess-enthalpy}) can be written as
\begin{equation}
H_{\rm ex} = \frac{3}{2} c_{\rm ex} RT_{g},
\label{eq:excess-enthalpy2}
\end{equation}
where $c_{\rm ex} = \overline{C_{\rm ex}}/3R$.
The enthalpy difference $\Delta H_{gc}$ between the glass and crystal is given by the difference between $H_{m}$ and $H_{\rm ex}$, if the width $\Delta T_{g}$ is ignored. It is then rewritten as
\begin{equation}
\Delta H_{gc} = 
\frac{3}{2} \left(1 - c_{\rm ex} \right) R T_{g}.
\label{eq:dHgc}
\end{equation}
The enthalpy difference $\Delta H_{gc}$ can vary from 0 to $H_{m}$, depending on $c_{\rm ex}$. Normally, there is a non-negligible difference in enthalpy between the glass and crystal. The size of $\Delta H_{gc}$ may be the same order as $R T_{g}$.
The enthalpy difference $\Delta H_{\rm tr} = H(T_{g2}) - H(T_{g1})$ at the transition region (called the transition enthalpy) is approximately given by
\begin{equation}
\Delta H_{\rm tr} \approx \frac{C_{p}^{(g)}(T_{g1})+C_{p}^{(l)}(T_{g2}) }{2} \Delta T_{g}
  = \frac{c_{\rm tr}}{k} R T_{g}.
\label{eq:transH}
\end{equation}
Here $c_{\rm tr}$ is the average specific heat in the transition region normalized by $R$.
The factor $k$ is the ratio of the transition temperature to its width, $k=T_{g}/\Delta T_{g}$. This factor plays an important role in the following arguments for characterizing the glass transition.\cite{Shirai21-ActEnergy} For fragile glasses, the value of $k$ is large, ranging from 10 to 40. Hence, the transition enthalpy $\Delta H_{\rm tr}$ is much less than $R T_{g}$.

Supercooling is caused by the intervention of an energy barrier at the boundary between solid and liquid parts of the system, which renders the transition irreversible.\cite{Shewmon} 
As is in the crystal nucleation, the energy barrier $E_{b}$ for the glass transition scales by the temperature difference $T_{m} - T_{g} \approx (1/2) T_{g}$. In the present case of glycerol, for example, a magnitude of about 10 meV is expected for $E_{b}$. 
Here is the long-term unsolved problem that the apparent activation energy $Q_{a}^{\ast}$ experimentally obtained is very much at variance with the expectation. The experimental value of glycerol is of the order of 1 eV: for example, 1.1 eV from viscosity measurement \cite{Davies53}, 1.8 eV from the measurement of enthalpy relaxation\cite{McMillan65}. These values are larger than the latent heat $H_{m}=14$ meV by two order of magnitude (note that $H_{m}$ is expressed per atom). In fact, unexpectedly large values for $Q_{a}^{\ast}$ are commonly observed for glasses. This general trend is formulated by an empirical relation
\begin{equation}
Q_{a}^{\ast} = 40 R T_{g},
\label{eq:Hunt}
\end{equation}
by Hunt.\cite{Hunt96} A similar but more amenable relationship is the one 
\begin{equation}
Q_{a}^{\ast} = 4.8 k RT_{g},
\label{eq:Moynihan}
\end{equation}
which is found by Moynihan from a series of inorganic glasses.\cite{Moynihan95} These two relations agree each other when $k$ is about 10.
All the above empirical relations support that the apparent activation energies $Q_{a}^{\ast}$ of glasses are  of the order of eV. It is indeed difficult to understand such unphysically large values $Q_{a}^{\ast}$ of glasses, unless we suppose that something goes wrong behind the conventional analysis of the Arrhenius plot, from which $Q_{a}^{\ast}$ is obtained. This was suggested by Goldstein.\cite{Goldstein10} The problem was solved by the first author of this paper.\cite{Shirai21-ActEnergy} 
When the structure is changed in a narrow range of temperature, $\partial \ln \eta/\partial (1/T)$ gives the apparent activation energy $Q_{a}^{\ast}$ magnified by the factor $k$: $Q_{a}^{*} = k E_{b}$. The factor $k$ can be looked upon the magnification factor for the energy barrier. For glycerol, since $T_{g}$ is reported to be from 180 to 190 K, the width is estimated at most to be $\Delta T_{g}=10$ K, resulting in $k=20$. Then the energy barrier is estimated as $E_{b} \approx 50$ meV, which falls in a reasonable range. By considering this magnification factor $k$ into account, the size of the real energy barrier $E_{b}$ can be estimated as
\begin{equation}
E_{b} = 4.8 RT_{g},
\label{eq:Eg-Tg}
\end{equation}
when the Moynihan's relationship (\ref{eq:Moynihan}) is used.
By comparing to Eq.~(\ref{eq:excess-enthalpy2}), we see that the energy barrier $E_{b}$ is the same order of magnitude as the excess enthalpy $H_{\rm ex}$. 
In summarize, from Eqs.~(\ref{eq:Hm-RTm}) to (\ref{eq:transH}) together with (\ref{eq:Eg-Tg}), the following relations for the size of energy are obtained, 
\begin{equation}
\Delta H_{\rm tr} \ll ( \Delta H_{gc} \sim H_{\rm ex} ) < H_{m} < E_{b}.
\label{eq:size}
\end{equation}
In the bracket, there is no rule as to which energy is larger. All the energies except $\Delta H_{\rm tr}$ are still of the same order of magnitude.

\subsubsection{Fragility}
\label{sec:fragility}
Finding the physical meaning of fragility is an important issue in glass physics, which stimulates interests of many researchers.\cite{Angell76, Takahara95,Ngai99,Johari00,Xia00,Martinez01,Wang-LM02, Wang03,Lubchenko03}
Now that the phonon contribution is excluded from the origin of $\Delta C_{p}$, the explanation of the empirical relation between dynamic and thermodynamic fragilities must be sought to the structural origin.
Fragility---more specifically kinetic fragility---is determined by the viscosity measurement. Although different ways of definition were proposed, the most appropriate one for the present purpose is the definition
\begin{equation}
m = \left( \frac{\partial \ln \eta}{\partial \ln T} \right)_{T=T_{g}},
\label{eq:fragility}
\end{equation}
which is also called the steepness index.\cite{Wang-LM02} This index $m$ represents the ratio $Q_{a}^{\ast}/RT_{g}$, as noted above. Angell noticed a general tendency that the stronger the fragility is the larger $\Delta C_{p}$ is.\cite{Angell95} A problem is how this qualitative tendency is formulated in a quantitative manner.
Martinez and Angell defined the thermodynamic fragility in terms of the excess entropy $S_{\rm ex}$: the steepness in the relation of $S_{\rm ex}(T_{g})/S_{\rm ex}(T)$ versus $T_{g}/T$.\cite{Martinez01} On defining this way, they found a good correlation between the kinetic and thermodynamic fragilities.\cite{Martinez01} An outstanding exception is silica glass, the reason for which is yet unknown.
A rather easy definition of the thermodynamic fragility is $\Delta C_{p}$ itself. Wang {\it et al.}~showed a quantitative relationship between the kinetic and thermodynamic fragilities by using this simple definition.\cite{Wang03, Wang06} They deduced the relationship,
\begin{equation}
m = a \frac{T_{g} \Delta C_{p}}{H_{m}},
\label{eq:m-delta_Cp}
\end{equation}
by analyzing experimental data, where $a$ is a constant. According to the relationship (\ref{eq:m-delta_Cp}), silica glass seems to fit the general tendency. Wolynes and his collaborators demonstrated that this relationship can be derived from the random first-order transition theory.\cite{Xia00,Lubchenko03} The constant $a$ in Eq.~(\ref{eq:m-delta_Cp}) is 56 in experiment \cite{Wang03, Wang06} while is 36 in theory \cite{Lubchenko03}. The agreement between theory and experiment, even for the constant $a$, confirms this relationship.

A more direct way to represent the relationship is use of the normalized specific heat change, $\Delta \tilde{C}_{p}(T)$, in the transition region.
By using the East-like models, Keys {\it et al.}~found a correlation between $\tilde{C}_{p}$ and fragility $m$ as, \cite{Keys13}
\begin{equation}
\frac{d \tilde{C}_{p}}{d \ln T} \approx 0.23 \ m.
\label{eq:Keys}
\end{equation}
By noting that $\tilde{C}_{p}$ varies from 0 to 1 in a narrow range $\Delta T_{g}$ of temperature, this relation can be rewritten as
\begin{equation}
\frac{T_{g}}{\Delta T_{g}} \approx 0.23 \ m.
\label{eq:Keys1}
\end{equation}
Alternatively, this is expressed as $m = 4.3 k$. 
The relation $m=Q_{a}^{\ast}/RT_{g}$ is interpreted that the fragility $m$ represents the $T$ dependence of the apparent activation energy.\cite{Dyre04} 
This interpretation is qualitatively correct. However, as stated above, the energy barrier $E_{b}$ is exaggareted to appear in the Arrhenius plot by the magnification factor $k$.\cite{Shirai21-ActEnergy} By the Moynihan's relation (\ref{eq:Moynihan}), the following relationship is derived,
\begin{equation}
m=4.8 k.
\label{eq:m-k}
\end{equation} 
This is close to the above relation~(\ref{eq:Keys}).
Thus we see that {\em the jump of specific heat at $T_{g}$ is a direct consequence of the steep change in the energy barrier in the transition region}. The amount of the energy change $\Delta H_{\rm tr}$ itself is small.

The Keys' equation (\ref{eq:Keys}) contains only the normalized specific heat $\tilde{C}_{p}$, and it is independent of the magnitude of $\Delta {C}_{p}$. In contrast, the Wang's equation (\ref{eq:m-delta_Cp}) is different in the point containing the magnitude $\Delta {C}_{p}$. In order to examine this relation, we have to extend our consideration to the $T$ region of supercooled liquid, $T_{g} < T < T_{m}$. 
The contributions from phonon and thermal expansion to $\Delta C_{p}$ in insignificant, so that $\Delta C_{p} = \Delta C_{\rm st}$. Furthermore, $C_{\rm st}^{(g)}$ can be approximately neglected compared with $C_{\rm st}^{(l)}$. By assuming linearity in $H^{(l)}$ with respect to $T$, we have
\begin{equation}
\Delta C_{p} \approx C_{p}^{(l)} = \frac{H_{\rm ex}}{T_{m}-T_{g}} = 2 \frac{H_{\rm ex}}{T_{g}},
\label{eq:Cp-liq}
\end{equation} 
Furthermore, $H_{m}$ is approximately proportional to the factor $k$, which is shown below. An energy barrier for $j$th atom is the energy change along a path of atom movement, $E_{b,j} = \int (\partial E/\partial R_{j})dR_{j}$ from its equilibrium position to the saddle point of atom migration with the distance $l_{j}$. The averaged energy barrier is approximately $E_{b} = \langle (\partial E/\partial R_{j}) l_{j} \rangle \approx \langle (\partial E/\partial R) \rangle \bar{l}$.
Since a small $H_{\rm ex}$ means that the structural difference between the glass and liquid is small, it is naturally expected that the energy barrier $E_{b}$ between the glass and liquid is smaller. The factor $k$ indicates how large the change in $E_{b}$ is when $T$ is changed. Hence, the energy difference $H_{\rm ex}$ between the glass and liquid is also scaled by $k$, 
\begin{equation}
H_{\rm ex} = b k H_{m},
\label{eq:HexHmk}
\end{equation} 
where $b$ is a constant. From Eqs.~(\ref{eq:m-k})-(\ref{eq:HexHmk}), the Wang's relationship (\ref{eq:m-delta_Cp}) is established, with the constant $a=4.8 b/2$ 
The value of $b$ can be estimated by investigating the following two extrema. For the case of no difference between the glass and liquid, $H_{\rm ex} = 0$ and hence $k=0$ is expected. On the other hand, for the case of no difference between the glass and crystal, $H_{\rm ex} = H_{m}$ is expected, which corresponds to that $k$ reaches the maximum value $k_{m}$. By assuming the maximum value $k_{m}=50$, $b=1/50$, leading to $a=120$. The difference in $a$ from the above values 56 or 36 is not so important, but the significance of Eq.~(\ref{eq:m-delta_Cp}) lies in that both quantities are scaled by the steepness of transition $k$. 
{\em The energy barrier most strongly changes where the internal energy changes}, as seen in Fig.~\ref{fig:HeatCycleGT}.
The former controls the kinetic properties of glass, while the latter determines thermodynamic properties.

%%%%%%%%%%%%%%%%%%%%%%%%%%%%%%%%%%%%%%%%%%
\section{Summary}
To calculate the specific-heat jump $\Delta C_{p}$ at $T_{g}$ without empirical parameters is an important step for understanding the glass transition. For this end, calculation of specific heat for general states of materials, regardless of solid or liquid, is indispensable. This paper presents a general method to calculate specific heat of materials. The method consists of two principles. The first is the total-energy approach based on DFT. Only the total energy is a reliable energy for general states of matter. Correspondingly, only the total specific heat is firmly grounded, while decomposition into parts is somewhat arbitrary. The second is use of adiabatic MD simulation. For liquid states, in particular, the transition state, the energy dissipations have an important contribution to the specific heat. The relationship between the internal energy and the temperature after reaching equilibrium contains all the effects of energy dissipation processes. 
Because no empirical relation is used, the present method has wide applicability to material sciences, such as specific heat at $\lambda$ transition.

An application of the present method has brought about fruitful results on the glass transition, through study on glycerol. The glass transition is simulated by a series of adiabatic MD runs by gradually changing the internal energy.
The small size of the supercell severely restricts the accuracy of calculation of $T_{g}$ and other thermodynamic quantities. In spite of this restriction, we have obtained relatively good agreement in the specific-heat jump $\Delta C_{p}$ for glycerol. The major part of $\Delta C_{p}$ comes from the change in the structural energy, while the contribution of softening of phonons at $T_{g}$ is insignificant. The latent heat of melting $H_{m}$ and the excess enthalpy $H_{\rm ex}$ are of the same order of magnitudes. The energy barrier $E_{b}$ of glass transition must be also the same order of magnitude. This energy scheme is compatible with the previous study that the apparent activation energy overestimates by the factor $k=T_{g}/\Delta T_{g}$. Finally, it is demonstrated that the relationship between the kinetic fragility and the specific-heat jump is naturally deduced from related energetics at the glass transition. The two quantities both scale with the the magnification factor $k$. Some relationships that exist in the literature have been explained in a consistent manner, which seem otherwise to be unrelated.
% despite their different appearances.
% ostensibly

%%%%%%%%%%%%%%%%%%%%%%%%%%%%%%%%%%%%%%%%%%
\section*{Acknowledgment}
The authors thank Kang Kim (Osaka Univ.)~for discussion on MD simulations of glasses. They also thank Y. Morikawa and I. Hamada (Osaka Univ.) for discussion on the calculation methods.
% Kang Kim
% We also thanks Myu Research (www.myu-inc.jp) for the English language review. 
We received financial support from the Research Program of ``Five-star Alliance" in ``NJRC Mater.~\& Dev."

%%%%%%%%%%%%%%%%%%%%%%%%%%%%%%%%%%%%%%%%%%
\bibliography{thermo-refs,glass-refs}

% \end{thebibliography}

%%%%%%%%%%%%%%%%%%%%%%%%%%%%%%%%%%%%%%%%%%
\end{document}